\documentclass[a4paper,10pt]{article}
\usepackage[utf8]{inputenc}
\usepackage{graphicx}

\newcommand{\be}{\begin{equation}}
\newcommand{\ee}{\end{equation}}
\newcommand{\bea}{\begin{eqnarray}}
\newcommand{\eea}{\end{eqnarray}}

\newcommand{\nn}{\nonumber}

\newcommand{\om}{\omega}

\newcommand{\oS}{\overline{S}}
\newcommand{\oSg}{\overline{\Sigma}}

\newcommand{\ov}{\overline}

\newcommand{\tom}{\widetilde{\om}}   

\newcommand{\vk}{\vec k}
\newcommand{\vp}{\vec p}

\newcommand{\vu}{\vec u}

\newcommand{\ks}{k \!\!\! /}

\newcommand{\ls}{l \!\!\! /}

\newcommand{\mn}{{\mu\nu}}
\newcommand{\del}{\partial}

\begin{document}

\title{A comparative analysis of in-medium spectral functions for $N(940)$ and $N^*(1535)$
in real-time thermal field theory}
\author{Sabyasachi Ghosh$^*$}
\date{}
\maketitle
\begin{center}
{$^*$Instituto de Fisica Teorica, Universidade Estadual Paulista, 
Rua Dr. Bento Teobaldo Ferraz, 271, 01140-070 Sao Paulo, SP, Brazil}
\end{center}

\begin{abstract}
In the real-time thermal field theory, the nucleon self-energy
at finite temperature and density is evaluated where an extensive
set of pion-baryon ($\pi B$) loops are considered. 
On the other side the in-medium self-energy of $N^*(1535)$
for $\pi N$ and $\eta N$ loops is also determined in the same
framework. The detail branch cut structures for these different 
$\pi B$ loops for nucleon $N(940)$ and $\pi N$, $\eta N$ loops for 
$N^*(1535)$ are addressed. Using the total self-energy of $N(940)$
and $N^*(1535)$, which contain the contributions of their corresponding 
loop diagrams, the complete structures of their in-medium spectral
functions have been obtained. The Landau and unitary cut contributions
provide two separate peak structures in the nucleon spectral function
while $N^*(1535)$ has single peak structure in its unitary cuts.
At high temperature, the peak structures of both at their individual poles are attenuated
while at high density Landau peak structure of nucleon is completely suppressed
and its unitary peak structure is tending to be shifted towards the 
melted peak of $N^*(1535)$. The non-trivial modifications of these
chiral partners may indicate some association of chiral symmetry restoration.
\end{abstract}


\maketitle

\section{Introduction}
In the limit of massless quarks ($u,~d$),
QCD Lagrangian density for quark field $\psi_f$ can be expressed
in terms of its left and right handed components 
$\psi^{L,R}_f=\frac{1}{2}(1\mp \gamma_5)\psi_f$ as
\bea
{\cal L}_{\rm QCD}&=&i\sum_{f=u,d}{\ov\psi}_f\gamma^\mu \del_\mu\psi_f+..
\nn\\
&=&i\sum_{f=u,d}{\ov\psi}^R_f\gamma^\mu \del_\mu\psi^R_f 
+i\sum_{f}{\ov\psi}^L_f\gamma^\mu \del_\mu\psi^L_f +..,
\label{mq_0}
\eea
which remains invariant under global SU$(2)_L\times$SU$(2)_R$ symmetry 
and leads to the conserved Noether currents 
$J^{\mu a}_{L,R}={\ov\psi}^{L,R}_f\gamma^\mu \frac{\tau^a}{2}\psi^{L,R}_f$. 
This implies that chirality or handedness is preserved 
and the associated symmetry of the strong interaction in this limit
is known as chiral symmetry. 

The observable particles i.e. hadrons are eigenstates of parity 
and so it is useful to work with the
vector and axial-vector Noether currents
\bea
J^{\mu a}_{V,A}&=&J^{\mu a}_{R} \pm J^{\mu a}_{L}
\nn\\
&=&{\ov\psi}_f\gamma^\mu\left\{ \begin{array}{ll}
\displaystyle {\bf 1}
\\
\displaystyle \gamma^5 
\end{array} 
\right\}\frac{\tau^a}{2}\psi_f~.
\eea
The triplet of charges $Q^a_{V,A}=\int d^3x J^{0a}_{V,A}(x)$ 
are the corresponding (quantum) generators of 
SU$(2)_R\times$SU$(2)_L$ which commute with the Hamiltonian 
of QCD
\be
[Q^a_{V,A},H]=0.
\label{H_Q_A}
\ee 
The states that from irreducible representation (basis) of the
SU(2)$_V$ group can be connected by 
\be 
Q_{V}|V_1\rangle =|V_2\rangle~.
\label{5_112}
\ee
From Eq.~(\ref{5_112}) and (\ref{H_Q_A}) it follows immediately that
\bea
E_{V_1} &=&\langle V_1|H|V_1\rangle
\nn\\
&=&\langle V_1|Q_V^\dagger H Q_V|V_1\rangle
\nn\\
&=&\langle V_2|H|V_2\rangle~~=E_{V_2}~.
\label{5_113}
\eea
Thus the symmetry of Hamiltonian $H$ is manifest in the degeneracies
of the energy eigenstates corresponding to the irreducible 
representations of the symmetry group. 
Since $|V_1\rangle$
and $|V_2\rangle$ must be related to the ground state $|0\rangle$
through some appropriate creation operators $\phi_{V_1}$ and $\phi_{V_2}$
by the relations: $|V_1\rangle=\phi_{V_1}|0\rangle$, $|V_2\rangle=\phi_{V_2}|0\rangle$
and $Q_V\phi_{V_1}Q_V^\dagger=\phi_{V_2}$. On the basis of these relations,
the Eq.~(\ref{5_112}) as well as Eq.~(\ref{5_113}) are satisfied 
only when~\cite{Cheng_Li} $Q_V|0\rangle=0$,
which was shown by Vafa and Witten~\cite{Witten}.
Isospin symmetry i.e. SU$(2)_V$ is consequently realized in 
the usual Wigner-Weyl mode which is reflected in the 
spectrum through the almost degenerate doublet of the 
proton and neutron, the triplet of the $\rho^+, \rho^0, \rho^-$ etc.
In addition to the vector charges, if the axial charges also 
annihilate the vacuum i.e. $Q^a_A|0\rangle=0$, parity doublets 
like scalar and pseudo scalar mesons ($\sigma,\pi$) or vector and axial
vector mesons ($\rho$, $a_1$) should exist in the spectrum. 
Although the vacuum hadronic spectra exhibit the absence of such 
kind of the doublets, which indicates $Q^a_A|0\rangle\neq 0$
by associating with the non-zero QCD vacuum, 
$\langle 0|{\ov \psi}_f\psi_f|0\rangle \neq 0$. These non-zero
relations lead to {\it spontaneous} or {\it Dynamical} breaking of chiral
symmetry (SBCS or DBCS)~\cite{Hayano,Holstein,Koch} even in the zero
quark mass limit, which prevent to break chiral symmetry {\it explicitly}.
Now under the extreme scenario of QCD matter produced
in heavy ion experiments at very high energy, this
broken symmetry may be restored by melting down
the quark condensate. Being associated with this chiral symmetry restoration (CSR),
the non-degenerate spectra of
chiral partners ($\sigma,\pi$), ($\rho$, $a_1$) etc. 
may approach towards the degenerate states under such an extreme
state of QCD matter.

In the baryon sector, an equivalent scenario is expected for nucleon
and its (lowest possible) chiral partner $N^*(1535)$.
In this context, a comparative investigation of in-medium spectral function
for $N(940)$ and $N^*(1535)$ may be very relevant and interesting.
This phenomenology of the baryons are analyzed by various 
groups~\cite{Lee,DeTar,DeTar2,Gottlieb,Jido,Jido2,Shuryak,Gallas} 
in different theoretical ways such as linear sigma model~\cite{DeTar}, 
lattice QCD calculations~\cite{DeTar2,Gottlieb}, 
QCD sum-rule approach~\cite{Jido,Jido2},
instanton liquid model~\cite{Shuryak} etc.
Here this article is intended to investigate this phenomenology
via effective hadronic model, where the thermodynamical parts
are governed by the real-time formalism of thermal field theory.
According to the Refs.~\cite{Koch,Kapusta_Shuryak}, 
the CSR mechanism may be linked with 
the different possible spectral modifications of the chiral partners.
Our aim is to search which one is preferred or indicated 
(may be partially) by our hadronic model 
calculation at finite temperature?

Next in the formalism part, the expression of thermal propagators
as well as self-energies for $N(940)$ and $N^*(1535)$ are explicitly 
derived. In Sec. (3), the detailed numerical results are discussed
and at last section the intention of the article is summarized.
\section{Formalism}
\subsection{Propagators of $N(940)$ and $N^*(1535)$ in the medium}

We begin with the 11 component of the nucleon propagator 
in real-time thermal field theory (RTF),
\be
S_{11}^{(0)}(k)=(\ks + m_N)D_{11}^{(0)}(k)~,
\ee
where
\bea
D_{11}^{(0)}(k)&=&\frac{-1}{k^2-m_N^2+i\eta} -2\pi i
F_k(k_0)\delta(k^2-m_N^2)~,
\nn\\
&&~~~~~~~~~~{\rm with}~~ F_k(k_0)=n^+_k\theta(k_0)+
n^-_k\theta(-k_0)
\nn\\
&=&-\frac{1}{2\om_k}\left(\frac{1-n^+_k}{k_0-\om_k+i\eta}+
\frac{n^+_k}{k_0-\om_k-i\eta}
\right.\nn\\
&&\left.~~~~~ -\frac{1-n^-_k}{k_0+\om_k-i\eta}
-\frac{n^-_k}{k_0+\om_k+i\eta}\right)~.
\label{de11}
\eea
Here $n^{\pm}_k(\om_k)=1/\{e^{\beta(\om_k \mp \mu_N)}+1\}$
denote Fermi-Dirac distribution functions of
nucleon anti-nucleon respectively with energy $\om_k=\sqrt{\vk^2+m_N^2}$. 

With the help of the diagonalization technique this 11 component
of thermal propagator can be transformed to the diagonal element
~\cite{Kobes,Ghosh_thesis},
\be
\oS^{(0)}(k)=(\ks + m_N)\frac{-1}{k^2-m_N^2+i\eta}~,
\ee
which is exactly same with free vacuum propagator.
The Dyson equation
in terms of the diagonal elements can be represented as~\cite{Kobes,Ghosh_thesis}
\be
\oS=\oS^{(0)}-\oS^{(0)}\oSg_N~\oS~,
\ee
where $\oS(k)$ and $\oSg(k)$ are
the diagonal element of complete propagator and
self-energy of nucleon respectively.
Taking scalar part of particle propagation only, we get 
the simplified form of nucleon spectral function,
\bea
&&A_N(k,T,\mu_N)={\rm Im}\oS(k,T,\mu_N)
\nn\\
&&=\frac{-{\rm Im}\oSg_N(k,T,\mu_N)}{(k_0-\om_k-{\rm Re}\oSg_N(k,T,\mu_N))^2
+({\rm Im}\oSg_N(k,T,\mu_N))^2}~,
\nn\\
\label{A_N}
\eea
where $\oSg_N={\rm Re}\oSg_N+i{\rm Im}\oSg_N$.
The $N^*(1535)$ will also have a similar form of spectral
function where the nucleon self-energy $\oSg_N$ will be 
replaced by it's own self-energy, $\oSg_{N^*}$.
\begin{figure}
\begin{center}
\includegraphics[scale=0.5]{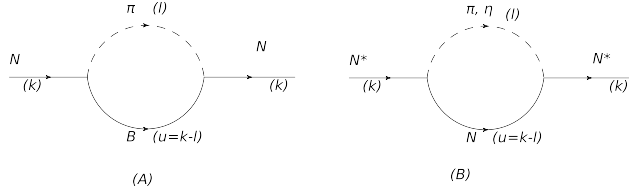}
\end{center}
\caption{Self-energy diagrams of $N(940)$ (A) and
$N^*(1535)$ (B) for respectively $\pi B$ and $\pi N$ (or $\eta N$)
loops.}
\label{NNst}
\end{figure} 
\subsection{self-energies of $N(940)$ and $N^*(1535)$ in the medium}
Next our aim is to calculate
the thermal self-energy of $N(940)$ and $N^*(1535)$
in RTF.
Inside a hot and dense nuclear matter,
the nucleon may be propagated via
different intermediate $\pi B$ loops, where $B$ stand for different
higher mass baryons including nucleon itself. An extensive
set of 4-star baryon resonances with spin one-half and three-half
are taken in this work. They are $N(980)$, $\Delta(1232)$, $N^*(1440)$, $N^*(1520)$,
$N^*(1535)$, $\Delta^*(1600)$, $\Delta^*(1620)$, $N^*(1650)$, 
$\Delta^*(1700)$, $N^*(1700)$, $N^*(1710)$, $N^*(1720)$;
where their masses (in MeV) are displayed inside the brackets.
The 
diagram~\ref{NNst}(A) has shown the nucleon self-energy for $\pi B$ loop.
The 11 component of self-energy can be represented as
\bea
\Sigma^{11}_N (k,T,\mu_N)&=&-i\int \frac{d^4l}{(2\pi)^4}L(k,l)D_{11}(l,m_\pi,T)
\nn\\
&&~~~~~~~D_{11}(u=k-l,m_B,T,\mu_N)~,
\label{Pi_11k}
\eea
where $D_{11}(l,m_\pi,T)$, $D_{11}(u=k-l,m_B,T,\mu_N)$ are scalar part of
the thermal propagators for pion and
baryon respectively. 
The two vertices and the numerator parts of the propagators
are contained in the factor $L(k,l)$.
All the baryon chemical potentials are supposed to be same
with nucleon chemical potential $\mu_N$.
Similar to the propagator matrix, the self-energy
matrix can also be diagonalized into a single component.
The diagonal element and
11 component are related as~\cite{Kobes,Ghosh_thesis}
\bea
{\rm Im}\oSg_N(k)&=&{\rm coth}\left\{\frac{\beta (k_0-\mu_N)}{2}
\right\}{\rm Im}\Sigma^{11}_N(k)
\nn\\
{\rm Re}\oSg_N(k)&=&{\rm Re}\Sigma^{11}_N(k)
\label{R_bar_N}
\eea
Taking the $l_0$ integration in (\ref{Pi_11k}) and then using
the relation (\ref{R_bar_N}), the imaginary and real part of 
the diagonal element can be obtained as
\bea
{\rm Im}{\oSg}_N(k)&=&\pi\int\frac{d^3l}{(2\pi)^3}\frac{1}{4\om_l\om_u}
\nn\\
&&[L_1\{(1+n_l-n^+_u)\delta(k_0 -\om_l-\om_u)
\nn\\
&&+(-n_l-n^-_u)\delta(k_0-\om_l+\om_u)\}
\nn\\
&&+L_2\{(n_l+n^+_u)\delta(k_0 +\om_l-\om_u)
\nn\\
&&+(-1-n_l+n^-_u)\delta(k_0 +\om_l+\om_u)\}]
\label{self_LU}
\eea
and
\bea
{\rm Re}{\oSg}_N(k)&=&{\cal P}\left[\int\frac{d^3l}{(2\pi)^3}\frac{1}{4\om_l\om_u}
\left\{\frac{L_1(1+n_l)-L_3n^+_u}{k_0 -\om_l-\om_u}
\right.\right.\nn\\
&&\left.\left.+\frac{-n_lL_1-n^-_uL_4}{k_0-\om_l+\om_u)}
+\frac{L_2n_l+L_3n^+_u}{k_0 +\om_l-\om_u}
\right.\right.\nn\\
&&\left.\left.
+\frac{-n_lL_2+(-1+n^-_u)L_4}{k_0 +\om_l+\om_u}\right\}\right]
\label{self_Re}
\eea
where $\om_l=\sqrt{{\vec l}^2+m_\pi^2}$, $\om_u=\sqrt{({\vk}-{\vec l})^2+m_B^2}$
and $L_i$ , $i = 1, . . . 4$ denote
the values of $L(l_0)$ for $l_0=\om_l,~-\om_l,~q_0-\om_l,~q_0+\om_l$
respectively.
The ${\cal P}$ indicates the principal value of the integrals. 
Here $n^\pm_u$ stand for Fermi-Dirac distribution functions 
of the baryons and anti-baryons 
while $n_l$ denotes the
Bose-Einstein distribution functions of the pion in the medium.
The range of 
the different branch cuts in $k_0$-axis are (
$-\infty$ to $-\{\vk^2+(m_\pi+m_B)^2\}^{1/2}$ ) for unitary cut in negative $k_0$-axis,
( $-\{\vk^2+(m_B-m_\pi)^2\}^{1/2}$ to $\{\vk^2+(m_B-m_\pi)^2\}^{1/2}$ ) for Landau cut and 
( $\{\vk^2+(m_\pi+m_B)^2\}^{1/2}$ to $\infty$ ) for unitary cut in positive $k_0$-axis.
Owing to the different $\delta$ functions in Eq.~(\ref{self_LU}),
the imaginary part of the nucleon self-energy become non-zero in the 
above regions.
We will mainly focus on the unitary and Landau cut contributions 
of Im$\oSg$ in positive $k_0$-axis, which are originated from
the first and third term of Eq.~(\ref{self_LU}) respectively. They
can be simplified as
\bea
{\rm Im}{\oSg}_N(k)&=&\frac{1}{16\pi\vk}[
\int^{\om^+_l}_{\om^-_l}d\om_l
L_1\{1+n_l(\om_l)-n^+_u(k_0-\om_l)\}
\nn\\
&&+\int^{\tom^-_l}_{\tom^+_l}d\tom_l
L_2\{n_l(\tom_l)+n^+_u(k_0+\tom_l)\}]
\label{gm_int}
\eea
where $\om^{\pm}_l=\frac{R^2}{2k^2}(k_0\pm\vk W)$,
$\tom^{\pm}_l=\frac{R^2}{2k^2}(-k_0\pm\vk W)$ with 
$W=\sqrt{1-\frac{4m_\pi^2k^2}{R^4}}$ and $R^2=k^2+m_\pi^2-m_B^2$.

The vacuum part of Re$\oSg_N$ is not written in the Eq.(\ref{self_Re})
as we are only interested in the medium part. That divergent quantity
is traditionally assumed to take part for generating physical mass
of nucleon.
\begin{table}   
\caption{From the left to right columns, the table contain
the baryons, their spin-parity quantum numbers $J_B^P$,
isospin $I_B$, total decay width $\Gamma_{\rm tot}$,
decay width in $N\pi$ channels $\Gamma_{B\rightarrow N\pi}$ or
$\Gamma_B(m_B)$ in Eq.~(\ref{Gam_BNpi}) (brackets displaying 
its Branching Ratio) and at the last coupling constants $f/m_\pi$.}
\label{tab}
\begin{tabular}{|c|c|c|c|c|c|}
\hline\noalign{\smallskip}
& & & & & \\
Baryons & $J_B^P$ & $I_B$ & $\Gamma_{\rm tot}$ & $\Gamma_{B\rightarrow N\pi}$ (B.R.) & $f/m_\pi$ \\
& & & & &  \\
\noalign{\smallskip}\hline\noalign{\smallskip}
& & & & & \\
$\Delta(1232)$ & ${\frac{3}{2}}^+$ & 3/2 & 0.117 & 0.117 (100\%) & 15.7 \\
& & & & & \\
$N^*(1440)$ & ${\frac{1}{2}}^+$ & 1/2 & 0.300 & 0.195 (65\%) & 2.5 \\
& & & & & \\
$N^*(1520)$ & ${\frac{3}{2}}^-$ & 1/2 & 0.115 & 0.069 (60\%) & 11.6 \\
& & & & & \\
$N^*(1535)$ & ${\frac{1}{2}}^-$ & 1/2 & 0.150 & 0.068 (45\%) & 1.14 \\
& & & & & \\
$\Delta^*(1600)$ & ${\frac{3}{2}}^+$ & 3/2 & 0.320 & 0.054 (17\%) & 3.4 \\
& & & & & \\
$\Delta^*(1620)$ & ${\frac{1}{2}}^-$ & 3/2 & 0.140 & 0.035 (25\%) & 1.22 \\
& & & & & \\
$N^*(1650)$ & ${\frac{1}{2}}^-$ & 1/2 & 0.150 & 0.105 (70\%) & 1.14 \\
& & & & & \\
$\Delta^*(1700)$ & ${\frac{3}{2}}^-$ & 3/2 & 0.300 & 0.045 (15\%) & 9.5 \\
& & & & & \\
$N^*(1700)$ & ${\frac{3}{2}}^-$ & 1/2 & 0.100 & 0.012 (12\%) & 2.8 \\
& & & & & \\
$N^*(1710)$ & ${\frac{1}{2}}^+$ & 1/2 & 0.100 & 0.012 (12\%) & 0.35 \\
& & & & & \\
$N^*(1720)$ & ${\frac{3}{2}}^+$ & 1/2 & 0.250 & 0.028 (11\%) & 1.18 \\
& & & & & \\
\noalign{\smallskip}\hline
\end{tabular}
\end{table}

The typical form of the $BN\pi$ interaction (effective) Lagrangian densities are as follows
~\cite{Leopold}
\bea
{\cal L}&=&\frac{f}{m_\pi}{\ov \psi}_B\gamma^\mu
\left\{
\begin{array}{c}
i\gamma^5 \\
1
\end{array}
\right\}
\psi_N\del_\mu\pi + {\rm h.c.}~{\rm for}~J_B^P=\frac{1}{2}^{\pm}~,
\nn\\
&=&\frac{f}{m_\pi}{\ov \psi}^\mu_B
\left\{
\begin{array}{c}
1 \\
i\gamma^5
\end{array}
\right\}
\psi_N\del_\mu\pi + {\rm h.c.}~{\rm for}~J_B^P=\frac{3}{2}^{\pm}~.
\label{Lag_BNpi}
\eea
The coupling constants $f/m_\pi$ for different $BN\pi$ 
interactions have been fixed from
the experimental vacuum widths of corresponding $B\rightarrow N\pi$
decays.
The free parameter of the Rarita-Schwinger field ($\psi^\mu_B$)
is chosen as $-1$~\cite{Peccie}.
Using the Lagrangian densities from Eq.~(\ref{Lag_BNpi}), one can easily
derive
\bea
L(k,l)&=&-\left(\frac{f}{m_\pi}\right)^2\ls(\ks-\ls -Pm_B)\ls
~~~~~~{\rm for}~J_B^P=\frac{1}{2}^{\pm}~,
\nn\\
&=&-\left(\frac{f}{m_\pi}\right)^2(\ks-\ls +Pm_B)l_\mu l_\nu
\left\{-g^{\mn}
\right.\nn\\
&&\left.+\frac{1}{3}\gamma^\mu\gamma^\nu
+\frac{2}{3m_B^2}(k-l)^\mu(k-l)^\nu
\right.\nn\\
&&\left.+\frac{1}{3m_B}(\gamma^\mu(k-l)^\nu-(k-l)^\mu\gamma^\nu)\right\}
\nn\\
&&~~~~~~~~~~~~~~~~~~~~~~~~~~~~~~~~{\rm for}~J_B^P=\frac{3}{2}^{\pm}~.
\eea
To avoid the complexity of dealing with Dirac structure of
self-energy as well as propagator, we have followed the simplified
technique of Ref.~\cite{Ghosh_N}, where total self-energy
has been identified as summation of the coefficients of
$\gamma^0$ and unit matrix.
Therefore, ignoring the coefficients of ${\gamma}^i$
for simplification and adding the coefficients of
$\gamma^0$ and unit matrix~\cite{Ghosh_N}, we have
\bea
L(k,l)&=&-\left(\frac{f}{m_\pi}\right)^2\left\{\left(k\cdot l-l^2
\right)l_0-Pl^2m_B\right\}
\nn\\
&&~~~~~~~~~~~~~~~~~~~~~~{\rm for}~J_B^P=\frac{1}{2}^{\pm}~,
\nn\\
&=&-\left(\frac{f}{m_\pi}\right)^2\frac{2}{3m_B^2}
\left\{\left(k\cdot l-l^2\right)^2
-l^2m_B^2\right\}(k_0
\nn\\
&&~~~~ -l_0+Pm_B)
~~{\rm for}~J_B^P=\frac{3}{2}^{\pm}~.
\eea
These vertex factors $L(k,l)$ have to be put in Eq.~(\ref{Pi_11k})
to obtain numerical values of nucleon self-energy.

The Lagrangian densities in (\ref{Lag_BNpi}) are not displaying
its isospin structures. 
For 
$J_B^P={\frac{1}{2}}^\pm$ and $J_B^P={\frac{3}{2}}^\pm$
these isospin structures should be ${\ov \psi}{\vec\tau}\cdot{\vec\pi}\psi$ and
${\ov \psi}{\vec T}\cdot{\vec\pi}\psi$ respectively, where
${\vec T}$ and ${\vec \tau}$ stand for the usual spin $3/2$ transition
and Pauli operator. These isospin structures provide 
appropriate isospin factors, which have to be multiplied with the expressions of 
corresponding $\pi B$ loop diagrams. 
The isospin factor for $\pi N$ or $\pi N^*$ loops is
$I_{N\rightarrow \pi N,N^*}=3$ and for the $\pi\Delta$
or $\pi\Delta^*$, it is $I_{N\rightarrow \pi \Delta,\Delta^*}=2$. 

Next we calculate vacuum width of different baryons in the $N\pi$
decay channel to fix their corresponding coupling constants $f/m_\pi$. 
With the help of the Lagrangian densities, vacuum
decay width of baryons $B$ for $N\pi$ channel can be
obtained as
\bea
\Gamma_B(m_B)&=&\frac{I_{N^*\rightarrow\pi N}}{2J_B+1}
\left(\frac{f}{m_\pi}\right)^2\frac{|\vp_{cm}|}{2\pi m_B}
[2m_B|\vp_{cm}|^2
\nn\\
&&+m_\pi^2(\om_N-Pm_N)]~~~{\rm for}~J_B^P=\frac{1}{2}^{\pm}~,
\nn\\
&=&\frac{I_{\Delta,\Delta^*\rightarrow\pi N}}{2J_B+1}
\left(\frac{f}{m_\pi}\right)^2\frac{|\vp_{cm}|^3}{3\pi m_B}
[\om_N
\nn\\
&&~~~~~~~+Pm_N]~{\rm for}~~~J_B^P=\frac{3}{2}^{\pm}
\label{Gam_BNpi}
\eea
where $|\vp_{cm}|=\frac{\sqrt{\{m_B^2-(m_N+m_\pi)^2\}\{m_B^2-(m_N-m_\pi)^2\}}}{2m_B}$
and $\om_N=\sqrt{|\vp_{cm}|^2+m_N^2}$.
The isospin factors are
$I_{N^*\rightarrow\pi N}=3$ and $I_{\Delta,\Delta^*\rightarrow\pi N}=1$
for the $N\pi$ decay channels of $N^*$ and $\Delta^*$ (or $\Delta$)
respectively.
Putting the experimental values~\cite{PDG} of $\Gamma_B(m_B)$ in Eq.~(\ref{Gam_BNpi}),
the values of coupling constants $f/m_\pi$ have been fixed, which are shown in
Table~(\ref{tab}).

For the self-energy calculation of $N^*(1535)$, the $\pi N$
and $\eta N$ loops are mainly considered because approximately
$45\%$ and $40\%$~\cite{PDG} of its vacuum width ($\Gamma_{N^*}=0.150$ GeV) 
are coming from these two decay channels ($\pi N$ and $\eta N$).
Hence the total self-energy of $N^*(1535)$ is defined as
\be
\oSg_{N^*}=\oSg^{\pi N}_{N^*}+\oSg^{\eta N}_{N^*}~,
\ee
where $\oSg^{\pi N}_{N^*}$ and $\oSg^{\eta N}_{N^*}$ are the
individual contributions for $\pi N$ and $\eta N$ loops.
They are diagrammatically shown in Fig.~\ref{NNst}(B).
In the imaginary part of self-energy for $T=0$, the remaining
part of vacuum width $0.022$ GeV ($15\%$ branching ratio)
are added with the numerical contributions of $\pi N$ and $\eta N$ loops.
The expressions of imaginary and real part of $\oSg^{\pi N}_{N^*}$
or $\oSg^{\eta N}_{N^*}$ will be similar with Eq.~(\ref{self_LU})
and (\ref{self_Re}) respectively, where
$\om_l=\{{\vec l}^2+m_{\pi,\eta}^2\}^{1/2}$ (for $\pi$ and $\eta$ respectively)and 
$\om_u=\{({\vk}-{\vec l})^2+m_N^2\}^{1/2}$ will be replaced only.  
Using Lagrangian density from (\ref{Lag_BNpi}) for $J_B^P=\frac{1}{2}^-$,
one can find the factor $L(k,l)$ for $\pi N$ loop as
\be
L(k,l)=-I_{N^*\rightarrow\pi N}\left(\frac{f}{m_\pi}\right)^2\left\{\left(k\cdot l-l^2
\right)l_0+l^2m_N\right\}~.
\ee
Again using the same Lagrangian density, where $\pi$ field only be replaced by
$\eta$ field, we can exactly receive same expression of $L(k,l)$ excluding
the isospin factor ($I_{N^*\rightarrow\pi N}=3$). The corresponding
coupling constant has also be replaced as it has been fixed
from the experimental decay width of $N^*(1535)$ in its $N\eta$ channel.

To include the in-medium effect of $N$ in the $\pi N$ or $\eta N$
loop, the modified $N^*(1535)$ self-energy can be defined as
(similar technique is used for $J/\psi$ in Ref.~\cite{Ghosh_D})
\bea
&&\oSg^{\pi,\eta N}_{N^*}(k,m_N,T,\mu_N)=\int_0^\infty dM^2 
\oSg^{\pi,\eta N}_{N^*}(k,M,T,\mu_N)
\nn\\
&&~~~~~~~~~~~~~~~~~~~~~
\left\{ \frac{A_N(u_0,\vu,T,\mu_N)}{\int_0^\infty dM^2A_N(u_0,\vu,T,\mu_N)} \right\}~,
\label{fold}
\eea
where $M^2=u_0^2-\vu^2$. This modified expression after folding
by nucleon spectral function at finite $T$ and $\mu_N$ can be restored
to its previous form (i.e. the form without folding) if the quantity inside $\{..\}$
of Eq.~(\ref{fold}) is replaced by $\delta(M^2-m_N^2)$.

\section{Results and discussion}

\begin{figure}
\begin{center}
\includegraphics[scale=0.35]{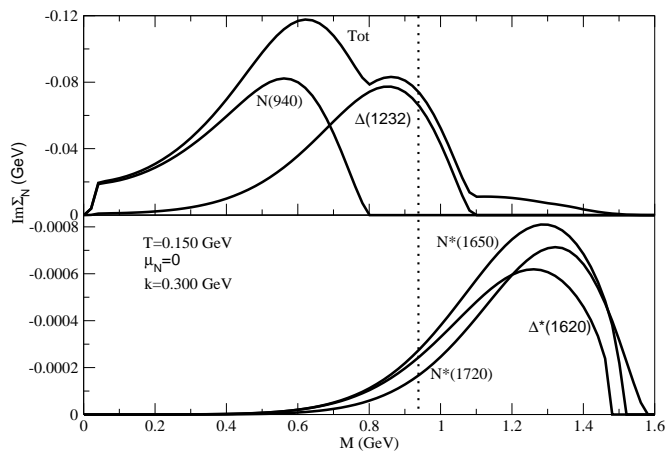}
\end{center}
\caption{Imaginary part of nucleon self-energy for different $\pi B$ 
loops. $B= \Delta^*(1620),~ N^*(1650),~N^*(1720)$ are shown in 
lower panel while $B= N(940),~\Delta(1232)$ and total of all loops 
are displayed in upper panel. }
\label{NIm_M}
\end{figure} 
\begin{figure}
\begin{center}
\includegraphics[scale=0.35]{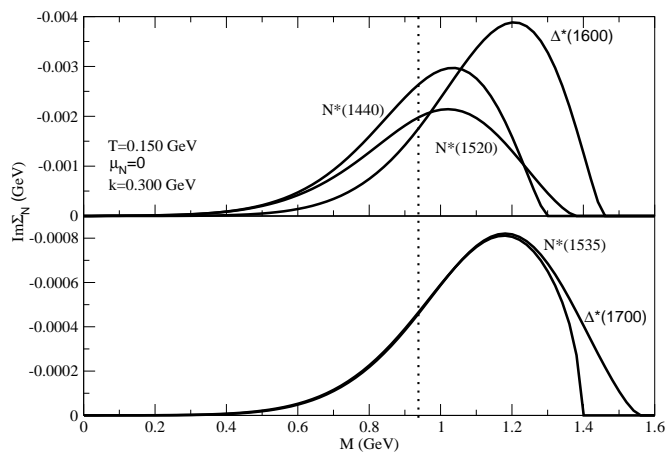}
\end{center}
\caption{Same as Fig.~(\ref{NIm_M}) for rest of the baryons
$B=N^*(1440),~ N^*(1520),~\Delta^*(1600)$ (upper panel) and
$B=N^*(1535),~\Delta^*(1700)$ (lower panel).}
\label{NIm_M2}
\end{figure} 

\begin{figure}
\begin{center}
\includegraphics[scale=0.35]{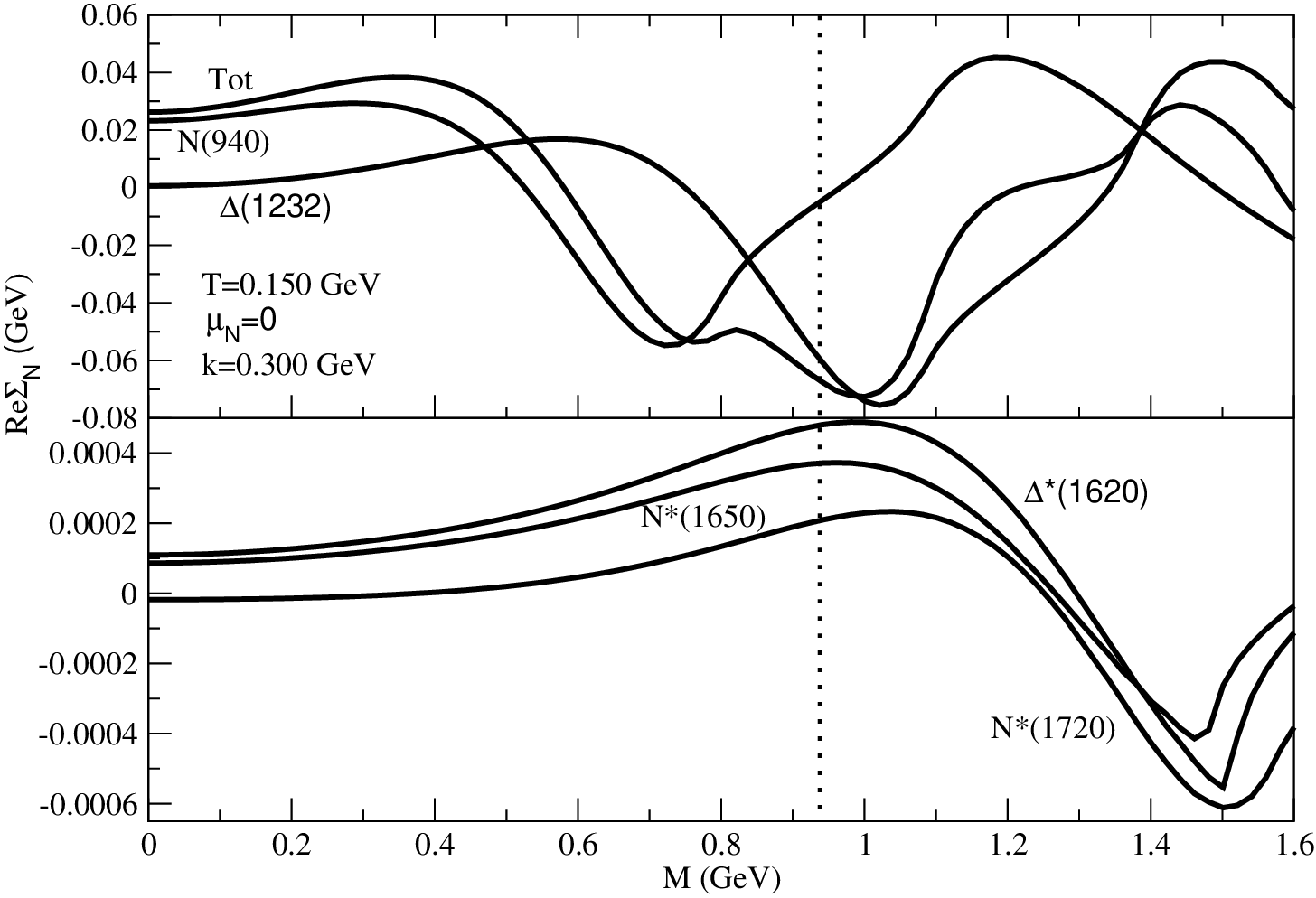}
\end{center}
\caption{The corresponding
results of Fig.~(\ref{NIm_M}) for the real part
of nucleon self-energy.}
\label{NRe_M}
\end{figure} 
\begin{figure}
\begin{center}
\includegraphics[scale=0.35]{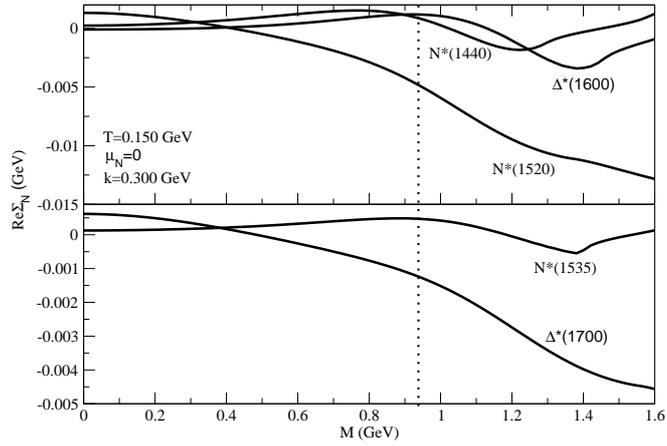}
\end{center}
\caption{The corresponding
results of Fig.~(\ref{NIm_M2}) for the real part
of nucleon self-energy.}
\label{NRe_M2}
\end{figure} 

\begin{figure}
\begin{center}
\includegraphics[scale=0.35]{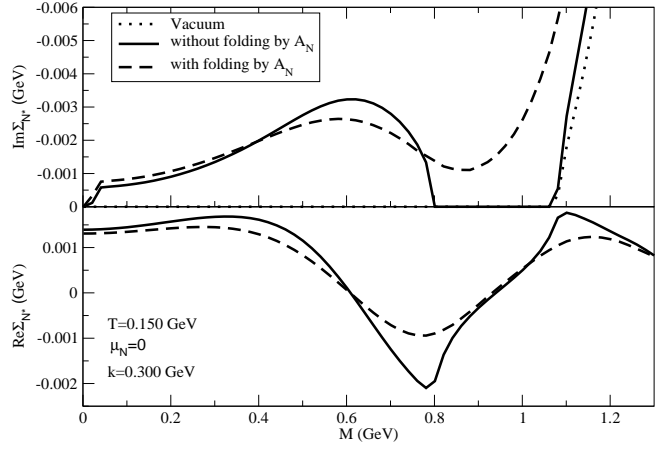}
\end{center}
\caption{Imaginary (upper panel) and real (lower panel)
part of $N^*(1535)$ self-energy  for $\pi N$ loop.
Dotted line exhibits the vacuum strength of Im$\Sigma_{N^*}$
coming from unitary cuts only.
The solid line of Im$\Sigma_{N^*}$ reveals two distinct regions 
of Landau and unitary cuts, which are overlapped after folding
Im$\Sigma_{N^*}$ by the in-medium nucleon spectral function, $A_N$.
This is shown in dashed line.}
\label{NstIm_M}
\end{figure}

\begin{figure}
\begin{center}
\includegraphics[scale=0.35]{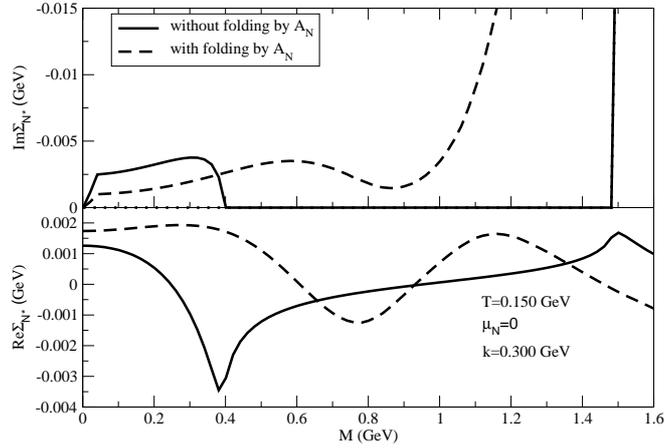}
\end{center}
\caption{Imaginary (upper panel) and real (lower panel)
part of  $N^*(1535)$ self-energy for $\eta N$ loop.
Same quantities of Fig.~(\ref{NstIm_M}) are represented
by solid and dashed lines for $\eta N$ loop.}
\label{NstIm_M2}
\end{figure}

\begin{figure}
\begin{center}
\includegraphics[scale=0.35]{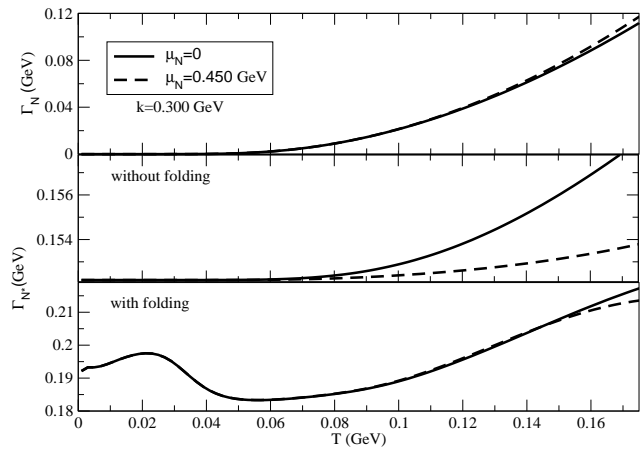}
\end{center}
\caption{The temperature dependence of $\Gamma_N$ (upper panel) 
and $\Gamma_{N^*}$ without (middle panel) and with (lower panel)
folding by the in-medium spectral function, $A_N$.}
\label{NNst_gm_T}
\end{figure}
Let us start with the results of different 
invariant mass
distribution of imaginary part of nucleon self-energy
for different $\pi B$ loops. 
In the Fig.~(\ref{NIm_M})
the results for baryons $B=N(940)$, $\Delta(1232)$ (upper panel)
and $B=\Delta^*(1620)$, $N^*(1650)$, $N^*(1720)$ (lower panel)
are shown whereas
the Fig.~(\ref{NIm_M2}) demonstrates the results for baryons 
$B=N^*(1440)$, $N^*(1520)$, $\Delta^*(1600)$ (upper panel)
and $B=N^*(1535), \Delta^*(1700)$ (lower panel). 
All these results are generated for fixed value of temperature ($T=0.150$ GeV),
nucleon chemical potential ($\mu_N=0$) and nucleon momentum ($\vk =0.300$ GeV).
The numerical strength for the $B=N^*(1700)$ and
$N^*(1710)$ are so low that they are not displayed with the other baryons.
The total contribution coming from all the loops is
displayed in the upper panel of Fig.~(\ref{NIm_M}).
The Landau regions
for different loops are clearly recognized 
from the sharp ending of solid lines for each loops. For example 
the Landau region of the $\pi N$ loop is ($M=0$ to $m_N-m_\pi$ i.e.
$0$ to $0.8$ GeV). 
The corresponding results of real part for different baryons are 
shown in Fig.~(\ref{NRe_M}) and (\ref{NRe_M2}). The contributions
imaginary or real part of nucleon self-energy at its pole have been
marked by dotted line in all of the graphs.

Similarly the imaginary (upper panel) and real (lower panel)
part of $N^*(1535)$ self-energy for $\pi N$ and $\eta N$ loops
are displayed in Figs.~(\ref{NstIm_M}) and (\ref{NstIm_M2}) respectively.
From the solid line of the figures (in the upper panel), the
Landau and unitary regions are distinctly observed. However,
their thresholds have been overlapped with each other after the 
folding by the in-medium spectral function of nucleon, $A_N$. 
The Eq.~(\ref{fold}) generates this with-folding results, which
are shown in dashed line in Fig.~(\ref{NstIm_M}) and (\ref{NstIm_M2}).
The thermal width $\Gamma_N$ for $N(940)$ and $\Gamma_{N^*}$
for $N^*(1535)$ are extracted from the pole contributions of
their corresponding total imaginary part of self-energy. For two
different values of $\mu_N$, the $T$ dependence of $\Gamma_N$
(upper panel), $\Gamma_{N^*}$ without (middle panel) and 
with (lower panel) folding are presented in Fig.~(\ref{NNst_gm_T}).
This non-zero $\Gamma_N(T,\mu_N=0)$ may have very important role
in different relevant quantities (e.g. in shear viscosity~\cite{Ghosh_Neta}),
estimated even for the baryon free matter, produced at RHIC or LHC experiments.
\begin{figure}
\begin{center}
\includegraphics[scale=0.35]{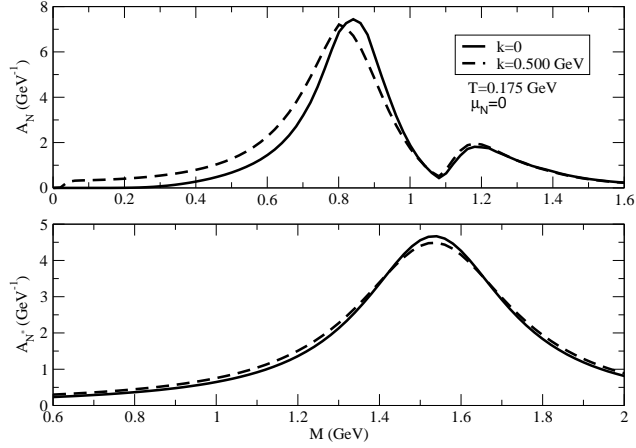}
\end{center}
\caption{The spectral function of $N(940)$ (upper panel) 
and $N^*(1535)$ (lower panel) 
for two different values of $\vk$.}
\label{spec_NNst_k}
\end{figure}
\begin{figure}
\begin{center}
\includegraphics[scale=0.35]{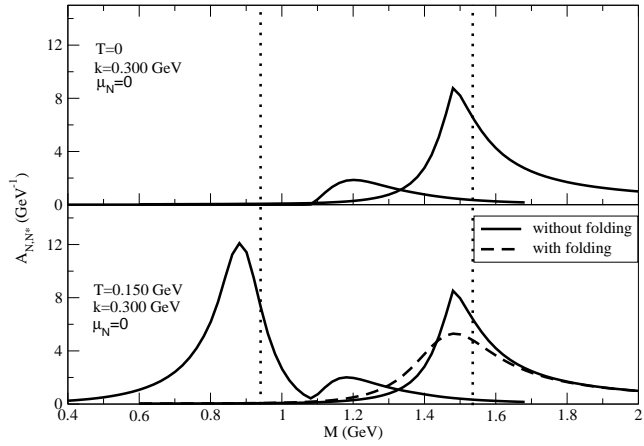}
\end{center}
\caption{The spectral function of $N(940)$ and $N^*(1535)$ 
for $T=0$ (upper panel) and $T=0.150$ GeV (lower panel).}
\label{spec_N_Nst}
\end{figure}
\begin{figure}
\begin{center}
\includegraphics[scale=0.35]{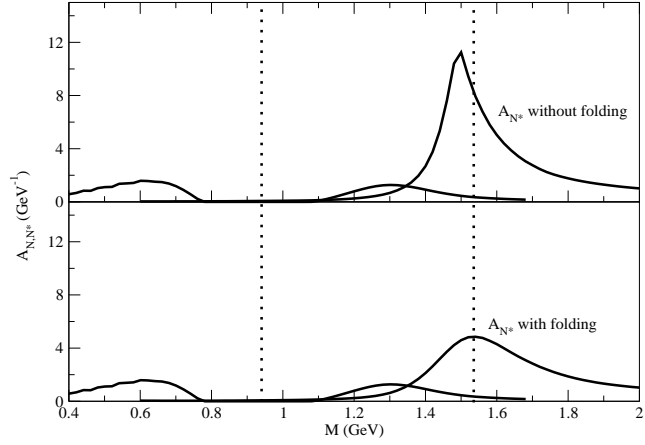}
\end{center}
\caption{The spectral function of $N(940)$ and $N^*(1535)$ 
for $T=0$, $\mu_N=0.976$ GeV and $\rho=\rho_0$. 
In the upper and lower panel, the results of $A_{N^*}$ without
and with folding (by $A_N$) are shown.}
\label{spec_NNst_mu}
\end{figure}
\begin{figure}
\begin{center}
\includegraphics[scale=0.35]{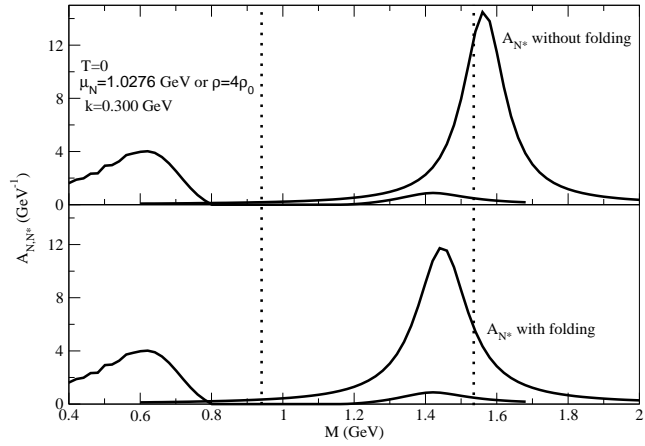}
\end{center}
\caption{The spectral function of $N(940)$ and $N^*(1535)$ 
for $T=0$, $\mu_N=1.0276$ GeV and $\rho=4\rho_0$.}
\label{spec_NNst_mu2}
\end{figure}

The total $N(940)$ self-energy contain the 
contributions of all of the $\pi B$ loops whereas the total
$N^*(1535)$ self-energy is composed of $\pi N$ and $\eta N$
loops. Using their total self-energy in the corresponding
expression of their spectral functions like
Eq.~(\ref{A_N}), we will get the explicit in-medium structure
of their spectral functions.  
The in-medium spectral functions of $N(940)$ (upper panel) and $N^*(1535)$
(lower panel) for two different values of three momentum $\vk$ are shown
in Fig.~(\ref{spec_NNst_k}). Unlike to vacuum case, the
spectral function at finite temperature become the function of
$k_0$ and $\vk$ independently which is numerically illustrated
in Fig.~(\ref{spec_NNst_k}).
The effect of temperature on their spectral functions is presented
in Fig.~(\ref{spec_N_Nst}) by displaying the results
for $T=0$ (upper panel) and $T=0.150$ GeV (lower panel).
As the threshold ($m_\pi+m_B$) of unitary cuts of the nucleon self-energy
is far away from the nucleon pole, a small magnitude of vacuum spectral 
function is obtained in the upper panel of Fig.~(\ref{spec_N_Nst}).
Since the nucleon pole is situated within the region of Landau cuts,
therefore at finite $T$, a good Breit-Wigner type structure is produced along with the 
small structures coming from the unitary 
cut contributions. This is displayed in lower panel of Fig.~(\ref{spec_N_Nst}).
Now for $N^*(1535)$ spectral function, the unitary cuts play 
a major role in vacuum as well as in medium as its pole is situated within
the unitary cuts for $\pi N$ and $\eta N$ loop diagrams. 
Due to folding by in-medium nucleon spectral function,
the thermal width of $N^*(1535)$ 
increases noticeably which 
is already shown in the lower panel of
Fig.~(\ref{NNst_gm_T}). Hence the peak structure of $N^*(1535)$ 
spectral function melts down after this folding which is denoted
by dash line in the lower panel of Fig.~(\ref{spec_N_Nst}).
The dotted lines are used to mark the bare pole positions
of $N(940)$ and $N^*(1535)$. 
At $T=0$ and $\mu_N=0.976$ GeV or $\rho=\rho_0$ 
(where $\rho_0=0.16$/fm$^3$ is the nuclear matter saturation density),
the modified spectral functions of $N(940)$ and $N^*(1535)$
are presented in Fig.~(\ref{spec_NNst_mu}) where the peak structure
of nucleon spectral function is completely suppressed. The peak 
of $N^*(1535)$ spectral function is slightly enhanced from its
vacuum spectral function because its thermal width will face
the Pauli suppression dominantly at $T=0$.
Although this peak strength has been slightly diminished after
the folding, which can be observed in the lower panel of 
Fig.~(\ref{spec_NNst_mu}). Again the peak of the small spectral 
strength for $N(940)$,
which is coming from its unitary cut contributions, is shifted
towards the peak of the $N^*(1535)$ spectral function. At very high
density their peak positions may be coincided with each other. 
This is shown in the lower panel of Fig.~(\ref{spec_NNst_mu2}).

Our first aim of this work is to investigate in-medium modifications
of $N(940)$ and $N^*(1535)$ spectral functions, whose vacuum strengths
are directly linked with the experimental inputs by 
fixing coupling constants of effective
Lagrangian densities. Analyzing the detailed branch cuts of their self-energies in RTF, 
their in-medium spectral profiles are exhibiting some non-trivial modifications.
Our next aim is to search any indication of CSR from their
non-trivial modifications. We should keep in mind that in the effective hadronic model,
the masses of chiral partners does not directly contain the information of temperature 
dependent quark condensate, which is nicely adopted in other chiral models like NJL,
LSM etc. Hence, a transparent indication of parity doublet may not be found as demonstrated
in the different chiral models. However, it is definitely an interesting to search
it in effective hadronic model approach, whose richness is the fixing of
interaction strengths via experimentally observed decay widths. During this searching of
link between CSR and these non-trivial modifications of $N(940)$ and $N^*(1535)$ spectral functions,
the conclusions of our hadronic model calculation in RTF are as follows. 
During increase of temperature, the Landau peak strength of $N(940)$, which
was completely absent in vacuum, may be approaching to be equal with the attenuated
unitary peak strength of $N^*(1535)$. Whereas, during increase of density (at $T=0$),
the Landau peak of $N(940)$ is hardly suppressed and its unitary
peak is shifted towards the unitary peak of $N^*(1535)$. Following the statements
mentioned in Refs.~\cite{Koch,Kapusta_Shuryak} regarding the possibilities spectral 
modifications of chiral partners to associate CSR, these approaching towards the equal
peak strength (at high temperature) and positions (at high density) of these chiral partners
may have some relation with CSR.

\section{Summary and conclusion}
To summarize, the in-medium self-energy of nucleon and it's 
chiral partner $N^*(1535)$ are evaluated in the RTF. An extensive
set of pion-baryon loops are taken for the nucleon self-energy
calculation. On the other hand $\pi N$ and $\eta N$ loops 
are considered for the $N^*(1535)$.  
After summing all the respective loop contributions for
$N(940)$ and $N^*(1535)$, their total self-energies have been
determined which provide them the complete structures of 
in-medium spectral functions. Two distinct
peak structure in the nucleon spectral function
have been originated from it's Landau and unitary cut contributions
whereas $N^*(1535)$ acquires a single peak structure from 
its unitary cuts. 
At high temperature, the spectral profile of both
are broadened with their attenuated peak structures.
At high density and $T=0$, 
the peak structure of nucleon spectral function,
coming from the Landau cuts, is completely suppressed
from its pole position. Along with this suppression of 
the Landau peak, the unitary peak structure is tending to be 
shifted towards the attenuated peak of $N^*(1535)$. 
This comparative investigation of in-medium spectral
functions for nucleon and its 
chiral partner $N^*(1535)$ exhibit 
a non-trivial modifications which may indicate
some association of chiral symmetry restoration.

{\bf Acknowledgment :} 
The work is financially supported
by Fundacao de Amparo a Pesquisa do Estado de Sao Paulo, 
FAPESP (Brazilian agencies) under Contract No. 2012/16766-0.
I am very grateful to Prof. Gastao Krein for his academic
and non-academic support during my postdoctoral period in Brazil.

\end{document}